# Measurement of neutron total cross-sections of $^{209}$Bi at the Pohang Neutron Facility*

WANG Tao-Feng (王涛峰) [1,2,♦], A.K.M.M.H.Meaze[3], Guinyun KIM[4]

[1] International Research Center for Nuclei and Particles in the Cosmos, Beihang University, Beijing 100191, China
[2] School of Physics and Nuclear Energy Engineering, Beihang University, Beijing 100191, China
[3] Department of physics, University of Chittagong, Chittagong 4331, Bangladesh
[4] Department of Physics, Kyungpook National University, Daegu 702-701, Korea

**Abstract**

Measurements of neutron total cross-sections of natural bismuth in the neutron energy region from 0.1 eV to 100 eV have been performed by using the time-of-flight method at the Pohang Neutron Facility, which consists of an electron linear accelerator, a water-cooled tantalum target with a water moderator, and a 12-m-long time-of-flight path. A $^{6}$Li-ZnS(Ag) scintillator with a diameter of 12.5 cm and a thickness of 1.6 cm is employed as a neutron detector, and a piece of high purity natural bismuth metallic plates with a thickness of 3 mm is used for the neutron transmission measurement. The present results were compared with the evaluated data from ENDF/B VII.1 and other previous reported experimental data.

**Key words**: Neutron total cross-section, Natural bismuth, Time-of-flight method, Pohang neutron facility
**PACS**: 28.20.Fc, 25.40.Ny, 24.30.Gd

## 1. Introduction

The cross-sections of neutron-induced reactions are basic nuclear data which could contribute to the safety design of nuclear reactors, to the medical application as well as to the development of nuclear waste transmutation system. Moreover, they are employed to the fundamental nuclear physics and the correlative research, for instance, the understanding of stellar nucleosynthesis. The lead-bismuth eutectic alloy coolant (LBC) used in the research fields of fast reactors and accelerator-driven system has attached a great deal of attention [1], since it has many advantages such as the characteristics of chemical inertness, high boiling point, low neutron moderation and large scattering cross section in comparison with the sodium coolant. $^{210}$Po is produced in LBC via the neutron capture reaction of $^{209}$Bi and the following β decay: $^{209}$Bi(n, γ)$^{210g}$Bi(β$^{-}$)$^{210}$Po, however, $^{210}$Po is an α-ray emitter and has strong radio-toxicity. Besides, $^{209}$Bi is the end-point element of s-process nucleosynthesis in stars, therefore the exact neutron cross sections are significant in the field of astrophysics.

Although some quantities of the advertent isotopes have already been measured previously, probably, the uncertainty and/or the energy range which has been measured are not satisfied with the requirements. There are a few transmission measurements of bismuth have been performed in the different neutron energy region. Yu. A. Alexandrov et al. [2] measured the neutron total cross sections of Bi with the conventional transmission method by a $^3$He ionization chamber on a linear accelerator in the neutron energy range from 0.1 to 90 eV. R. E. Mayer et al. [3] employed seven $^3$He proportional counters to perform a measurement for Bi from 0.1 to 582 eV with time of flight (TOF) method. A. B. Popov et al. [4] made an experiment for total neutron cross-section of bismuth on a fast pulse reactor in connection with iron, molybdenum and cobalt filters. Besides, J. A. Harvey et al [5]. measured the same quantities by using a $^6$Li glass scintillator with 1.27cm thickness on an electron linear accelerator at ORELA.

*Supported by the National Natural Science Foundation of China under Grant No 11305007 and 11235002)
♦ E-mail: tfwang@buaa.edu.cn
This paper has already been submitted to Chinese Physics C

In the present work, neutron transmission spectroscopy for natural bismuth sample have been measured by the neutron TOF method at Pohang Neutron Facility (PNF) [6] which consists of an electron linear accelerator, a water-cooled tantalum target with a water moderator, a 12-m long time-of-flight path and a $^6$Li-ZnS(Ag) scintillator with 3 mm thickness as neutron detector. This paper reports the neutron total cross-sections of natural Bi in the neutron energy region from 0.1 eV to 100 eV. The present results were compared with the evaluated data from ENDF/B VII.1 [7] and experimental data from literatures.

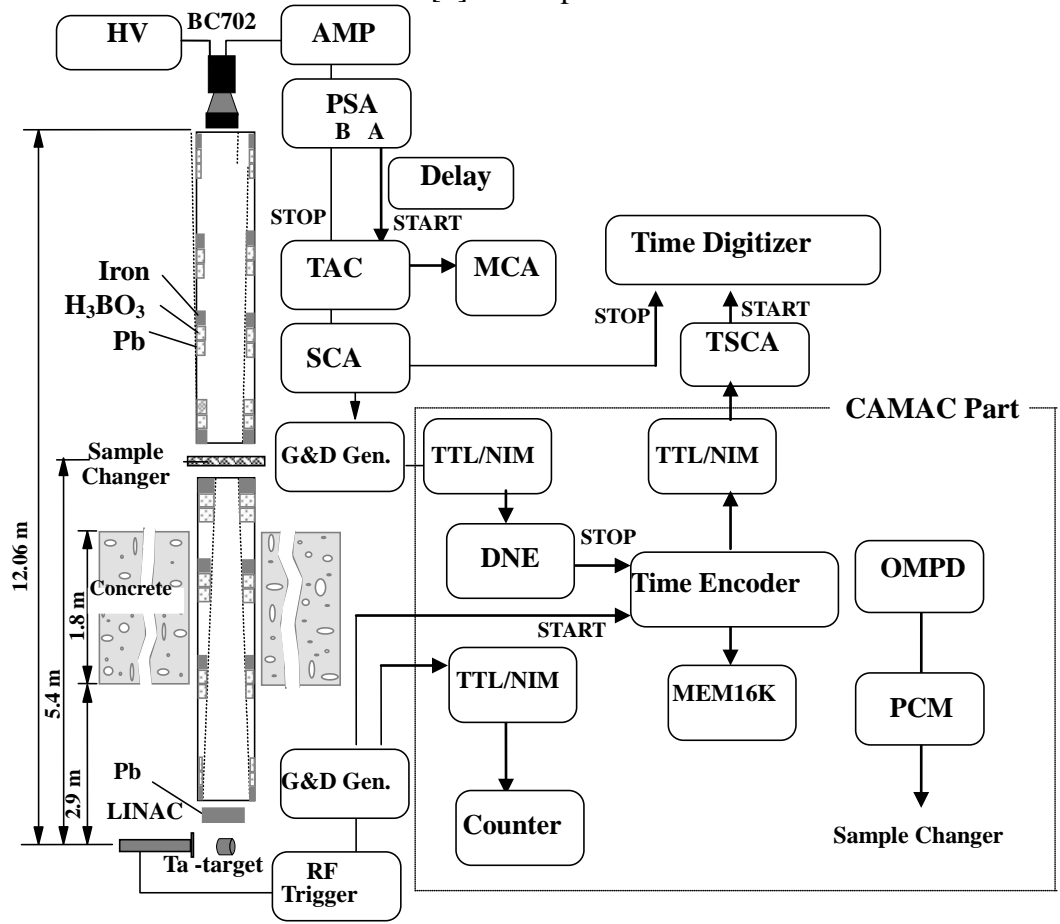

Fig.1. Configuration of experimental setup and data acquisition system.

**2. Experimental procedure**

2.1. Experimental arrangement

The neutron transmission measurement for natural bismuth samples were performed by time-of-flight method at the pulsed neutron source PNF, since the experimental procedure has been published previously [8], only a general description is given here. The experimental arrangement for the transmission measurement is shown in Fig. 1. Pulsed neutrons were produced via the $^{181}$Ta (γ, xn) reaction by bombarding a metallic Ta-target with the pulsed electron beam. The estimated neutron yield per kW of beam power for electron energies above 50 MeV at the Ta target based on the calculation by MCNP code to be $1.9\times10^{12}$ n/s [9], which is consistent with the calculated value based on Swanson's formula, $1.21\times10^{11}$ $Z^{0.66}$, where Z is the atomic number of the target material [10]. To maximize the thermal neutrons in this facility, we have to use a cylindrical water moderator contained in an aluminum cylinder with a thickness of 0.5 cm, a diameter of 30 cm, and a height of 30 cm. The water level in this experiment was 3 cm above the target surface.

Pulsed neutron beam was collimated to 5-cm diameter in the middle position of collimation system where the sample changer was located. The sample changer was controlled remotely by using a CAMAC module. The neutron detector was located at a distance of about 12 m from the photo-neutron target. A $^6$Li-ZnS (Ag) scintillator (BC702) with a diameter of 12.5 cm and a thickness of 1.6 cm mounted on an EMI-93090 photomultiplier was used as a neutron detector. During the transmission measurement, the electron linac was operated with a repetition rate of 15 Hz, a pulse width of 1.0 µs, and the electron energy of 70 MeV. The peak current in the beam current monitor located at the end of the second accelerator section was greater than 50 mA, which was almost the same as that in the target.

Table 1. Physical parameters of transmission samples.

| Samples | Size (mm) | Thickness (mm) | Purity(%) | Density (g/cm$^3$) |
|---|---|---|---|---|
| $^{209}$Bi | 50.06±0.01×50.14±0.01 | 3.0±0.1 | 99.9 | 9.8 |
| $^{59}$Co | 50.72±0.01×50.37±0.01 | 0.505±0.001 | 99.95 | 8.90 |
| $^{115}$In | 50.04±0.01×50.50.59±0.01 | 0.479±0.001 | 99.9 | 7.31 |
| Cd | 50.78±0.01×50.08±0.01 | 0.436±0.001 | 99.85 | 8.65 |

2.2 Samples

A high-purity (99.9%) bismuth metal plate shaped as 50-mm × 50-mm square sheets with the natural isotopic composition ($^{209}$Bi100%) was used as the transmission measurement sample. A set of notch filters of Co (purity 99.99%), In (purity 99.99%), and Cd (purity 99.99%) plates which have the same size as that of Bi sample were used for the background measurement and the energy calibration. The physical parameters of the samples used in this experiment are given in Table 1. During the transmission measurement samples were placed at the midpoint of the neutron flight path and were cycled into the neutron beam by using an automatic sample changer with four sample positions.

2.3 Data acquisition system and experiment

The configuration of the data acquisition system used in this measurement is also show in Fig.1. Two different data acquisition system were used for the neutron TOF spectra measurements: one for a NIM-based system and the other for a CAMAC-based system. The main purpose of the NIM-based system is neutron-gamma separation [11] and parallel accumulation of the neutron TOF spectra, if necessary. The Time Digitizer (150-MHz Turbo Multichannel Scaler) was operated by a 16384-channel mode with a 0.5 μs width per channel. The CAMAC-based system consists of a data acquisition part and a control part of the sample changer (SC). The SCA output signal of the NIM-based system is connected to the detector number encoder (DNE) through an NIM/TTL converter. The DNE allows the data to be taken from up to four detectors simultaneously. The output of the DNE is the stop signal for the time encoder which operates with 4096 channels and a minimal dwell time of 0.5-μs in width per channel.

The sample changer consists of a disc with 4 holes; each hole is 8-cm in diameter, which matched the hole of the collimator in the neutron beam line. The exposure time for the sample and open position of one cycle were set as 300 seconds during the whole measurement. This interleaving sequence of free positions of the sample changer was chosen to minimize the influence of slow and/or small variation of the neutron beam intensity. If the beam intensity variation or its drift was fast and/or large, then these partial measurements were excluded from the total statistics. The total data taking time for Bi sample was 48 hours, this data taking time was same as that of the open beam measurement.

## 3. Data analysis and results

The time-of-flight spectrum of neutron transmission for notch filter consisted of Co, In and Cd plates was shown in Fig. 2, the specific resonance dips in the spectrum used for background level estimation correspond to the energy positions of Cd (<0.025 eV), $^{115}$In (1.457 eV) and $^{59}$Co (132 eV), respectively. This energy range is same as that of transmission measurement for Bi samples. The magnitude of the background level was interpolated between the black resonances by using the fitting function $y = A\ exp(\ I/t\ ) + B$, where $A$, $t$ and $B$ are constants and $I$ is the channel number of the time digitizer, as shown in Fig. 2.

We obtained the flight path length $L$ in meters from the resonance energy $E$ in eV corresponding to the channel number $I$ as indicated in Fig. 2 by using the following fitting function,

$$I = \frac{72.3 \times L}{\Delta W \times \sqrt{E}} + \frac{\tau}{\Delta W} \tag{1}$$

$$E = \left[\frac{72.3 \times L}{I \times \Delta W - \tau}\right]^2 \tag{2}$$

where $\Delta W$ is the channel width of the time digitizer that we used 0.5 μs and $\tau$ is the time difference between the start time from the RF trigger and the real time zero when the neutron burst was produced. The neutron flight path length $L$ was determined by fitting the resonance channel numbers of Co and In as a function of its corresponding neutron energy using Eq. (1). We found $L = 12.06 \pm 0.02$ m and $\tau = 7.15 \pm 0.01$ μs from the fitting procedure.

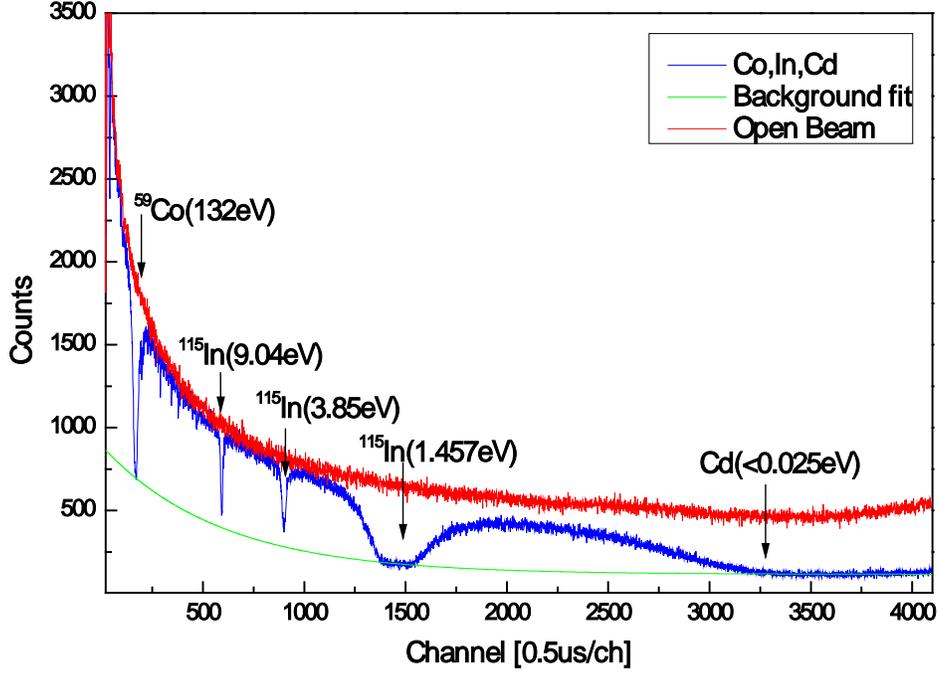

Fig. 2. Background level determination with Co, In and Cd samples. The background fitting function was determined using the black resonance energy of the notch filters.

Then, the energy resolution can be written as

$$\frac{\Delta E}{E} = 2\frac{\Delta t}{t} \qquad (3)$$

where the uncertainty ($\Delta t$) of the neutron TOF ($t$) is composed of uncertainties due to the flight path (2 cm), the moderator thickness (3 cm), the pulse width of the electron beam (1 μs), the channel width of the time encoder (0.5 μs), and the time jitter (negligibly small) from the neutron detector. The energy resolutions for the neutron energy of 0.01, 0.1, 1, 10, 100 and 200 eV are 0.59%, 0.60%, 0.65%, 1.01%, 2.63% and 3.68 %, respectively.

The accumulated neutron TOF spectrum of open beam operations and of the natural Bi sample are shown in Fig.3, together with the estimated background level, which is indicated by a solid line.

The transmission of an incident beam of neutrons at the $i$-th group energy $E_i$ on a sample with a thickness of $n$ atoms per barn is defined as the relation to the total cross section $\sigma_T(E)$ by the following expression

$$T(E_i) = e^{-n\sigma_T(E_i)} = \frac{\varphi_1(E_i)}{\varphi_2(E_i)} = N\frac{S_{in}(E_i) - B_{in}(E_i)}{S_{out}(E_i) - B_{out}(E_i)} \qquad (4)$$

$$\sigma_T(E_i) = -\frac{1}{n}\ln T(E_i) \qquad (5)$$

where the ratio of attenuated and incident neutron fluxes $\varphi_1(E_i)$ and $\varphi_2(E_i)$ is named as the transmission factor $T(E_i)$, $S(E_i)$ is time-integrated counts corrected for the estimated background $B(E_i)$ for the situation of sample in and sample out of the neutron beam, $N$ is the normalization factor, in this measurement, we assumed the monitor counts to be equal during the whole measurement process.

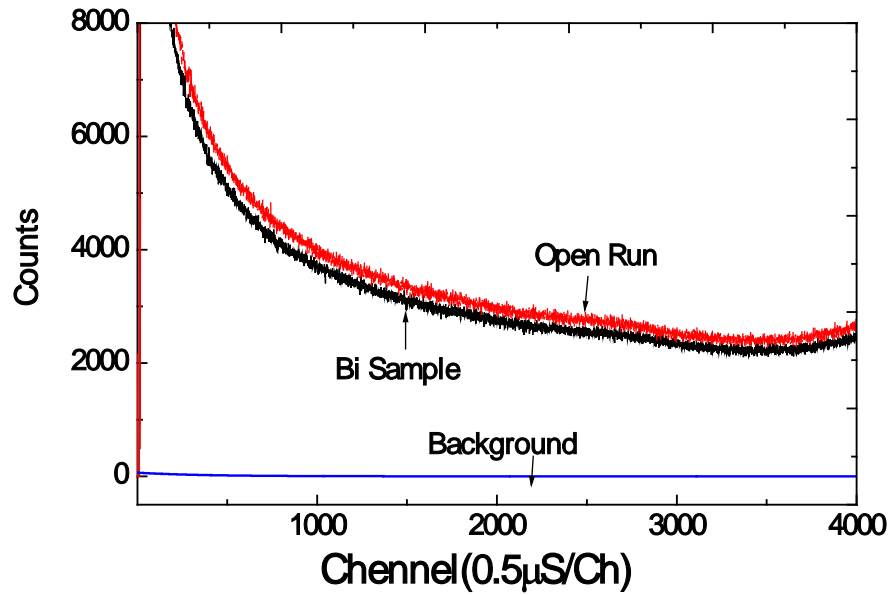

Fig. 3. Neutron time-of-flight spectra for Bi sample and for the open beam (sample out) together with the estimated background level.

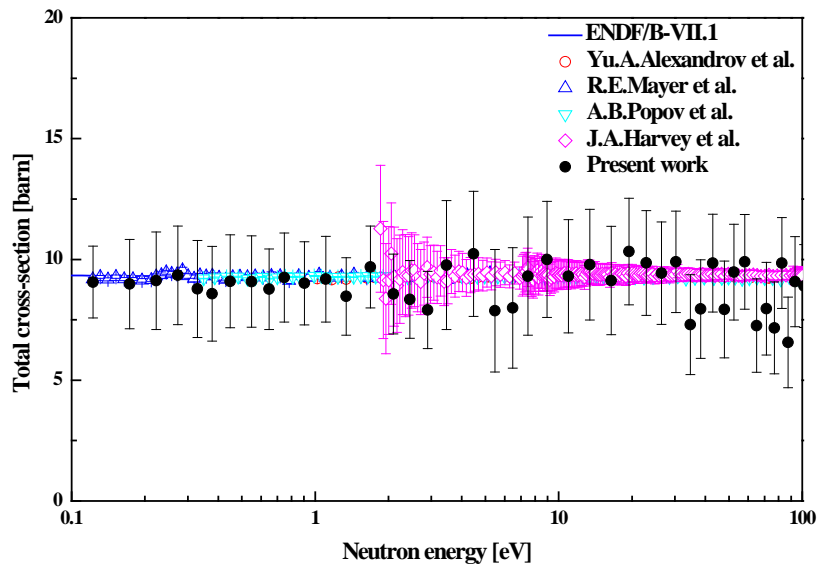

Fig. 4. The effective total cross sections of natural bismuth of a thickness 3 mm compared with previous experiment data and evaluation data from ENDF/B-VII.1 in the energy range from 0.18eV to 100eV.

After applying various spectrum manipulations on the recorded histograms mentioned above, energy dependence of effective total cross sections of natural bismuth obtained from Eq. (5) were compared with previous experiment data [2-5] and evaluation data from ENDF/B-VII.1, as shown in Fig. 4. The present results below 1 eV are in good agreement with the data of R. E. Mayer et al. [3] and Yu A. Alexandrov et al. [4]. Even though there are little fluctuations in the data between 2 to 30 eV, our results are almost consist with the work of J. A. Harvey et al. [5]. The values in present measurements between 30 to 100 eV are smaller compared to the results of J. A. Harvey et al. [5], the possible reason is that the neutron flux is lower with the increase of the incident neutron energy, besides, the bismuth sample is too thin for the low reaction cross sections, a higher statistic record data are needed.

**4. Conclusion**

We have measured the neutron total cross sections of natural Bi sample from 0.18 eV to 100 eV at the Pohang Neutron Facility by using the neutron TOF method and the $^6$Li–ZnS(Ag) scintillator as a neutron detector. The present results are compared with other measured values and the evaluated values in ENDF/BVII.0 [7]. The overall data are generally in good agreement with previous measurements and the evaluated data.

*The authors would like to express their sincere thanks to the staff of the Pohang Accelerator Laboratory for the excellent operation of the electron linac and their strong support.*

**References**
[1] Saito K, Igashira M, Kawakami J et al. J. Nucl. Sci. Technol.,2004, **41**:406
[2] Alexandrov Yu A, Guseva I S, Laptev A B et al. Int. Sem. On Interactions of Neutrons with Nuclei, 1997, **5**: 255, http://www-nds.iaea.org/EXFOR/41382.009
[3] Mayer R E, Gillette V H, Granada J R et al., http://www-nds.iaea.org/EXFOR/30747.003
[4] Popov A B, Samosvat G S, http://www-nds.iaea.org/EXFOR/40932.002
[5] Harvey J A, Good W M et al., http://www-nds.iaea.org/EXFOR/12816.002
[6] Kim G N, Lee Y S, Skoy V et al., J. Korean Phys. Soc., 2001, **38**: 14
[7] Chadwick M B, Herman M, Oblozinsky P et al. Nucler data sheets, 2011, **112**: 2887
[8] Wang T F, Meaze A K M M H, Khandaker M U et al., Nucl. Instr. Meth. B, 2008, **266:** 561
[9] Nguyen V D et al., J. Korean Phys. Soc.,2006, **48** : 382
[10] Swanson, William P., Health Phys. **35** (1978) 353.
[11] Kim G N, Ahmed H, Machrafi R et al, J. Korean Phys. Soc., 2003, **43**: 479
[12] Derrien H, Guber K H, Harvey J A et al, J. Nucl. Sci. Technol., 2002, Supplement 2, 84
[13] Goncalves I F, Martinho E, Salgado J, Nucl. Instr. Meth. B, 2004, **213**: 186